\documentclass[11pt,a4paper,svgnames]{article}
\pdfoutput=1

\usepackage[utf8]{inputenc}
\usepackage{uniinput}
\usepackage{jheppubmod}
\usepackage{hyperref}
\usepackage{amsmath}
\usepackage{url}
\usepackage{bm}
\usepackage{mathtools}
\usepackage{relsize}
\usepackage{cancel}
\usepackage{lipsum}
\newcommand{\D}{\mathrm{d}}
\newcommand{\ie}{\textit{i.e.~}}

\newcommand{\bea}{\begin{eqnarray}}
\newcommand{\eea}{\end{eqnarray}}
\newcommand{\be}{\begin{equation}}
\newcommand{\ee}{\end{equation}}
\usepackage[format=plain,
textfont=it
]{caption}
\usepackage{microtype}

\title{Dark bubbles: decorating the wall}
\author[a]{Souvik Banerjee,}
\emailAdd{souvik.banerjee@physik.uni-wuerzburg.de}
\affiliation[a]{Institut für Theoretische Physik und Astrophysik,
	Julius-Maximilians-Universität Würzburg,Am Hubland, 97074 Würzburg, Germany\\}
\author[b]{Ulf Danielsson,}
\emailAdd{ulf.danielsson@physics.uu.se}
\affiliation[b]{Institutionen för fysik och astronomi,
	Uppsala Universitet, Box 803, SE-751 08 Uppsala, Sweden\\}
\author[b]{Suvendu Giri}
\emailAdd{suvendu.giri@physics.uu.se}

\abstract{
Motivated by the difficulty of constructing de Sitter vacua in string theory, a new approach was proposed in \href{https://arxiv.org/abs/1807.01570}{arXiv:1807.01570} and \href{https://arxiv.org/abs/1907.04268}{arXiv:1907.04268}, where four dimensional de Sitter space was realized as the effective cosmology, with matter and radiation, on an expanding spherical bubble that mediates the decay of non supersymmetric AdS₅ to a more stable AdS₅ in string theory. In this third installment, we further expand on this scenario by considering the backreaction of matter in the bulk and on the brane in terms of how the brane bends. We compute the back reacted metric on the bent brane as well as in the five dimensional bulk. To further illuminate the effect of brane-bending, we compare our results with an explicit computation of the five dimensional graviton propagator using a holographic prescription. Finally we comment on a possible localization of four dimensional gravity in our model using two colliding branes.
}
\preprint{UUITP-1/20}

\begin{document}
    \maketitle
    \section{Introduction and overview}\label{sec:general}
    The extreme difficulty of constructing a model of the observed positive cosmological constant in string theory, has led to the recent proposal of de Sitter (dS) swampland conjectures \cite{Garg:2018reu,Obied:2018sgi,Ooguri:2018wrx}, which forbid the existence of such vacua in string theory (see \cite{Danielsson:2018ztv} for a review). To get around this difficulty, a possible new approach was proposed in \cite{Banerjee:2018qey,Banerjee:2019fzz}, where four dimensional dS is realized as a time-dependent geometry in an unstable five dimensional AdS vacuum of string theory. This scenario is inspired by the Randall-Sundrum braneworld \cite{Randall:1999ee,Randall:1999vf} construction, but differs from it in a crucial way.  Instead of a flat brane with a $\mathbb{Z}₂$ symmetry across it, our \textit{shellworld} is a spherical brane that mediates the decay of an unstable five dimensional AdS to a more stable one. This leads to important differences in the way that four dimensional matter and radiation is realized on the shellworld. For instance, a cloud of strings stretching out from the shell leads to matter, while a black hole in the bulk give rise to radiation on the four dimensional shellworld. In fact, it was shown in \cite{Koga:2019yzj} that the presence of these ingredients in the five dimensional bulk facilitate the decay of the unstable AdS and favor the formation of a bubble with a small cosmological constant.
    
    Let us begin by summarizing the shellworld construction of 
    \cite{Banerjee:2018qey,Banerjee:2019fzz}. Consider a theory that has two AdS₅ vacua with cosmological constants $Λ_+$ $(= -6k_+²=-6/L_+²)$ and $Λ_-$ $(= -6k_-²=-6/L_-²)$ with $k_->k_+$. The vacuum with higher energy $Λ_+$ can decay into the vacuum with lower energy $Λ_-$ non-perturbatively via the nucleation of a Brown-Teitelboim instanton. \footnote{In string theory, such decays are supported by the conjecture that all non supersymmetric AdS vacua must decay as a consequence of the weak gravity conjecture \cite{Ooguri:2016pdq,Danielsson:2016mtx,Danielsson:2017max}.}  The decay AdS vacua supported by flux through the nucleation of charged membranes, and its relation to the Weak Gravity Conjecture \cite{ArkaniHamed:2006dz} in the context of shellworlds, was discussed in \cite{Banerjee:2019fzz}.
    Let us consider these AdS vacua in global coordinates (where the subscripts $+$ and $-$ denote quantities outside and inside \footnote{Inside (outside) refers to the direction away from the brane in which the volume of radial slices decreases (increases).} the bubble respectively)
    \begin{equation}\label{eq:metric}
    \D s² = -f_\pm(r) \D t² + f_\pm(r)^{-1} \D r² + r² \D \Omega_3²,
    \end{equation}
    where $f_\pm(r) = 1+k_\pm ²r²$.
    For the metric inside the bubble with $f_-(r)$, the radial coordinate $r$ goes from the centre of AdS to the position of the bubble \ie $r \in \left(0,a(t)\right)$, while for the metric outside the bubble with $f_+(r)$, $r \in \left(a(t),∞\right)$. The complete metric for all $r$ \ie $r \in \left(0,∞\right)$ in the five dimensional bulk is then given by 
    \begin{equation}
    \D s² = -f(r)\D t²+f(r)^{-1}\D r²+r²\D \Omega_2²,
    \end{equation}
    with
    \begin{equation}
    f(r)\coloneqq (1+k_-²r²)+Θ(r-a(t))(k_+²r²-k_-²r²),
    \end{equation}
	
    On parametrizing the radius of the bubble in terms of the proper time ($τ$) for an observer on the shell \ie, $r=a(τ)$, the induced metric takes the FLRW form
    \begin{equation}
    \D s²_{\rm{ind}} = -\D τ² + a(τ)²\D \Omega_3².
    \end{equation}      
    Einstein's equations require the presence of a stress tensor on the shell given by
    \begin{equation}
    \label{eq:second-junction}
    S_{ab} = -\frac{1}{κ₅²}\left(\left[K_{ab}\right]^+_--\left[K\right]^+_-γ_{ab}\right),
    \end{equation}
    where $κ₅²\equiv 8πG₅$, \footnote{We follow the standard conventions where the coefficient of the five dimensional Ricci scalar in the Einstein Hilbert action is $M_5³=1/(2κ_5²)=1/(16πG₅)$, where $M_5$ is the five dimensional Planck mass.} and $γ_{ab}$ is the metric induced on the shell from the bulk, where the indices $(a,b)\in\left\{0,1,2,3\right\}$ run over the shell. $K_{ab}$ is the extrinsic curvature of the shell as seen from the AdS bulk, and $\left[\cdot\right]^+_- \coloneqq (\cdot)_+ - (\cdot)_-$ is the difference of the corresponding quantity between the outside and the inside of the bubble. 
    This governs the evolution of the radius of the bubble, and for a shell of constant tension $σ$ it is given by
    \begin{equation}\label{eq:tension}
    σ = \frac{3}{κ₅²}\left( \sqrt{k_-² + \frac{1+\dot{a}²}{a²}} - \sqrt{k_+² + \frac{1+\dot{a}²}{a²}} \right),
    \end{equation}
    where $a(τ)\equiv a$ and $\dot{a}\coloneqq \D a(τ)/\D τ$. For a critical value of the tension $σ_{\rm{crit}}$ given by
    \begin{equation}
    σ_{\rm{crit}} \coloneqq \frac{3}{κ₅²} \left(k_- - k_+\right),
    \end{equation}
    the cosmological constant on the shell vanishes, giving a four dimensional Minkowski brane. 
    For a brane with a slightly sub-critical tension $σ=σ_{\rm{crit}}(1-\epsilon)$, equation \eqref{eq:tension}, up to linear order in $\epsilon$, gives the Friedmann equation
    \begin{equation}
    H² \equiv \frac{\dot{a}²}{a²} = -\frac{1}{a²}+\frac{κ₄²}{3}Λ₄+\mathcal{O}\left(\epsilon ²\right),
    \end{equation}
    where
    \begin{align}
    \label{eq:G4}
    κ₄² \equiv 8πG₄ &\equiv \frac{2k_+k_-}{k_- - k_+}κ₅²,&
    Λ₄ &\equiv σ_{\rm{crit}} - σ = \epsilon σ_{\rm{crit}}.
    \end{align}
    Therefore, an observer living on such an expanding shell experiences a de Sitter universe with a positive spatial curvature. 
    
    In the presence of matter in the five dimensional AdS, the metric is given by equation \eqref{eq:metric} with $f_\pm(r)=1+k_\pm ²-κ₅²M_\pm/(3π²r²)$. This gives an additional contribution to the Friedmann equation which goes as $1/a^4$ and can be identified as a radiation density in four dimensions \ie,
    \begin{equation}
    H² = -\frac{1}{a²}+\frac{κ₄²}{3}\left[Λ₄ + \frac{1}{2π²a⁴}\left(\frac{M_+}{k_+}-\frac{M_-}{k_-}\right)\right].
    \end{equation}
    Four dimensional matter can be obtained from a cloud of strings in the bulk, which has $f_\pm(r)=1+k_\pm ²-κ₅²M_\pm/(3π²r²)-κ₅²α/(4πr)$, and gives the following contribution to the Friedmann equation
    \begin{equation}\label{eq:Friedmann}
    H² = -\frac{1}{a²}+\frac{κ₄²}{3}\left[Λ₄ + \frac{1}{2π²a⁴}\left(\frac{M_+}{k_+}-\frac{M_-}{k_-}\right)
    + \frac{3}{8πa³}\left(\frac{α_+}{k_+}-\frac{α_-}{k_-}\right)\right].
    \end{equation}
    In summary, a positively spatially curved four dimensional de Sitter universe can be modeled as the effective spacetime seen by a four dimensional observer living on the surface of an expanding shell in an asymptotically five dimensional AdS. The matter and radiation densities observed in the four dimensional universe are induced by the presence of matter and a cloud of strings in the five dimensional bulk.
    
    This model also provides a way of understanding the dS horizon in four dimensions \cite{Banerjee:2019fzz}.
    A four dimensional observer living on the shell accelerates in the fifth dimension with the expanding shell and measures an Unruh temperature. This turns out to be equal to the temperature associated with the dS horizon in four dimensions. The dS temperature thus has a five dimensional interpretation in terms of the acceleration of the expanding shell. This means that the cosmological horizon is a section of the Rindler horizon in five dimensions.
    
    Gravitational dynamics on the shell can be studied using the Gauss~(-Codazzi) equations, which relate the Einstein equations in the bulk to those on the shell in terms of its embedding and extrinsic curvature,
    \begin{equation}\label{eq:Gauss}
    R^{(5)}_{αβγδ} e^α_c e^β_a e^γ_d e^δ_b = R^{(4)}_{cadb} + (K_{ad}K_{cb}-K_{cd}K_{ab}).
    \end{equation}
    Combined with the thin shell junction conditions, it was shown in \cite{Banerjee:2019fzz} that the four dimensional Einstein equations are given by
    \begin{equation}\label{eq:T4}
    \begin{split}
    \left(G^{(4)}\right)^a_{b} =&- \underbrace{2k_+ k_-\left(3 -\frac{κ₅²}{k_--k_+}σ\right)}_{\equiv κ²_4\left(σ_\textrm{crit}-σ\right) \equiv  κ²_4Λ_4}
    δ^a_b 
    -\frac{κ²₅}{π²a(τ)⁴}\left(\frac{M_+k_- - M_-k_+}{k_--k_+}\right)
    \left(δ^a_0δ^0_b-\frac{1}{3}\sum_{i=1}^3δ^a_iδ^i_b\right)\\
    &-\frac{3κ²₅}{4πa(τ)³}\left(\frac{\alpha_+k_- - \alpha_-k_+}{k_--k_+}\right)δ^a_0δ^0_b
    %+2κ²_5\left(\frac{k_+ k_-}{k_- - k_+}\right)\left(T_\textrm{brane}\right)^a_b
    ,
    \end{split}
    \end{equation} 
    The left hand side of this equation is the Einstein tensor, and the right hand side gives the sources corresponding to the four dimensional cosmological constant, radiation and matter respectively. Each contribution is easily identified through its equation of state given by $p=-ρ,p=ρ/3$ and $p=0$, respectively. Taking the $(τ,τ)$ component of the Einstein tensor in four dimensions, namely $G^{(4)}_{ττ}=3\left(1/a²+\dot{a}²/a²\right)$ and using the above equation, gives the Friedmann equation \eqref{eq:Friedmann} as expected.                
    
    In the current paper, we will re-derive \eqref{eq:G4} through a computation of the graviton propagator in momentum space. In this computation, we will explicitly use intuitions coming from holography \cite{Susskind:1994vu}. Holography, in particular the AdS/CFT correspondence \cite{Maldacena:1997re}, relates a gravitational theory in $(d+1)$ dimensional anti de Sitter spacetime to a non-gravitational $d$-dimensional conformal field theory living on the boundary of the $(d+1)$ dimensional \textit{bulk} spacetime. In its full generality, both the bulk and boundary theories are quantum theories. However, in certain limits \footnote{namely, in the limit of large number of degrees of freedom, $N \rightarrow \infty$ in the boundary field theory and large t'Hooft coupling $\lambda = g_{\textrm{\sc qft}}^2 N \gg 1$, $g_{\textrm{\sc qft}}$ being the coupling in the boundary field theory.} the gravitational theory becomes classical as well as weakly coupled. In this limit, there is a well-defined prescription \cite{Witten:1998qj, Gubser:1998bc} that relates bulk quantities to boundary data. 
    In particular, the bulk modes that decay with radial distance near the boundary (termed as the \textit{normalizable modes}) are interpreted as the expectation value of the corresponding operator on the quantum field theory on the boundary, while the other independent set of relatively slowly growing or constant modes (namely, the \textit{non-normalizable modes}) have the interpretation of sources. 
    As an example, for the graviton propagator, normalizable modes correspond to the stress tensor while the non-normalizable modes correspond to the metric on the boundary.
    We will not get into much of the technical details of holographic prescription in this paper \footnote{Interested readers are referred to the elaborate review \cite{Aharony:1999ti} and references therein.}, but only use this intuition while fixing the constants of integration when computing the bulk propagator. 
    
    Our main focus in this paper will be the following. In the construction of \cite{Banerjee:2018qey,Banerjee:2019fzz}, the shell is radially symmetric and is located at a constant radius in the five dimensional bulk at each instant of time. The presence of four dimensional Einstein gravity on the brane was shown in the presence of homogeneous and isotropic matter. However, in the presence of localized matter on the brane, breaking homogeneity and isotropy, it is expected to bend \cite{Garriga:1999yh,Giddings:2000mu}.
%    In the presence of matter on the brane, it is expected to bend \cite{Garriga:1999yh,Giddings:2000mu}. 
    This backreaction, which has not been considered so far, is one of the main aspects that we will focus on in this paper. In addition, we will also compute the metric on the bent brane as well as the back-reacted five dimensional metric.
    
    \begin{figure}
    	\centering
    	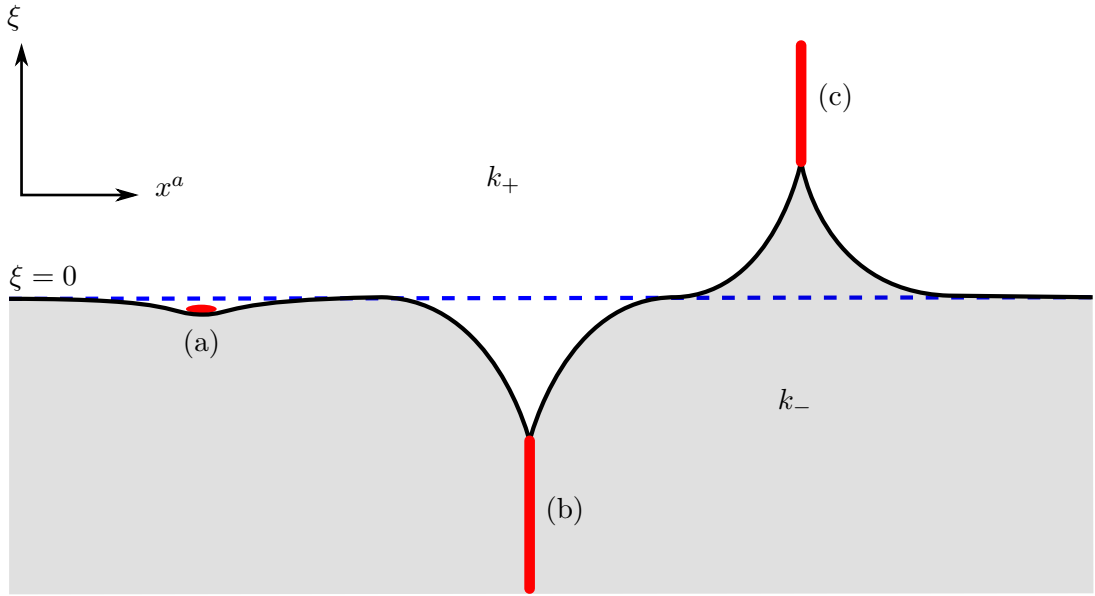
    	\caption{A schematic representation of the brane, showing the effect of various forms of stress tensor. The brane is shown in black and the sources are in red. (a) A point mass causes the brane to ``sag'' \ie, $ξ<0$. A gauge transformation to bring the brane back up to $ξ=0$ is physically equivalent to adding a negative four dimensional mass to the brane to ``float'' it back up. (b) A string pulling on the brane from the inside has the same effect, while (c) a string pulling from outside has the opposite effect and contributes with a positive four dimensional energy in a gauge where the brane is flat.}
    	\label{fig:brane_bending}
    \end{figure}
    
    In figure \ref{fig:brane_bending}, we summarize the contribution of matter sources to the bending of the brane. We will discuss this in detail in section \ref{sec:brane_bending} and here we present a geometrical understanding of the main result. 
    In the coordinates that we are using, the Poincaré horizon is located at $ξ→-∞$ and the boundary of AdS is at $ξ→∞$, so gravity pulls down towards the center of the AdS throat, implying that matter confined to the brane will make it sag (case (a) in figure \ref{fig:brane_bending}). The brane will bend in a similar way with a string pulling downwards (case (b) in figure \ref{fig:brane_bending}). As we will argue in subsequent sections, it is the case with strings pulling the brane upwards (case (c) in figure \ref{fig:brane_bending}) that will be the most interesting. The presence of such strings will affect the metric in the bulk and on the brane in two different ways. There is the change in the metric sourced directly by the string that is present even without the brane, as well as a contribution from the bending through the junction conditions. As we will see explicitly in section 3.2., the direct effect is subleading at distances much larger than the AdS-scale, where the effect from the bending dominates.
    
    In order to distinguish between the two effects, let us invoke the covariant conservation of the stress tensor \ie, $\nabla_αT^{αβ}=0$. \footnote{The Bianchi identities require the stress tensor to be covariantly conserved in order for Einstein's equations to be consistent. This is a constraint on the components of the stress tensor that we will use here to find the stress tensor of the end point of the string in terms of the tension of the string. This is similar to the treatment in \cite{Giddings:2000mu}.}  For $β=ξ$, the conservation condition gives (assuming no cross terms in $ξ$ \ie, $T^{aξ}=0$)
    \begin{equation}\label{eq:covT}
    ∂_ξ T^{ξξ} + 4kT^{ξξ}+ka(ξ)²η_{ab}T^{ab}=0.
    \end{equation}
    Recalling that the metric is $\D s² = \D ξ² + a(ξ)²η_{ab}\D x^a \D x^b$, a string of constant tension $τ$ stretching only in the $ξ$ direction (and located at $x^a=x^a_0$) has an energy density given by
    \begin{equation}
    T_0^0=T^ξ_ξ=τ\frac{1}{a(ξ)³}δ\left(x^a-x^a_{0}\right).
    \end{equation}
    This corresponds to an equation of state $p=ρ$ along the fifth direction and $p=0$ in the transverse direction,
    which is exactly as expected for a string stretching out infinitely only in the $ξ$ direction. If, instead, the string ends on the brane (located at $ξ=ξ_0$, say), then $T_ξ^ξ$  becomes
    \begin{equation}
    T_ξ^ξ = τ\frac{1}{a(ξ)³}δ\left(x^a-x^a_{0}\right)Θ\left(ξ-ξ_0\right).
    \end{equation}
    Equation \eqref{eq:covT} then gives
    \begin{equation}
    T_0^0=\underbrace{τ\frac{1}{a(ξ)³}δ\left(x^a-x^a_{0}\right)Θ\left(ξ-ξ_0\right)}_{\textrm{string stretching in the bulk}}
    +\underbrace{\frac{τ}{k}\frac{1}{a(ξ)³}δ\left(x^a-x^a_{0}\right)δ\left(ξ-ξ_0\right)}_{\textrm{end point on the brane}}.
    \end{equation}
    The first term is the expected energy density of the string stretching in the bulk, but now there is a second contribution on the brane corresponding to the point where the string ends on the brane. This is the energy density of a point particle with mass $τ/k$ exactly as was argued from the Friedmann equations in \cite{Banerjee:2019fzz}. It is useful to think of this as a point mass 
    hanging by a string in the gravitational well of the AdS-space. As the universe expands, and the brane moves upwards, the string pulls the point mass so that it can move along with the brane.
    
    This is the setup that we will be revisiting throughout this paper. The organization is as follows. In section \ref{sec:brane_bending}, we will introduce the idea of brane bending in the presence of matter sources. We will start with a leading order computation demonstrating the effect of brane bending and then perform a more detailed analysis to show which way the brane bends in response to being pulled by strings in the bulk. In section \ref{sec:lineargrav}, we will compute the graviton propagator first in momentum space and then in position space. Subsequently, we will discuss how these computations are related by a gauge choice. Finally, in section \ref{sec:conclusion}, we will highlight some future directions. In particular, we will discuss the possibility of fully localizing gravity in the shellworld model using two approaching bubbles, the possibility of realizing black holes on the shellworld and a realization of holography on a cut-off brane.
    
    \section{Brane bending: a primer}\label{sec:brane_bending}
    \noindent
    In this section, we will study the effect of \textit{brane bending} in response to matter (both in the bulk and on the brane) that we briefly introduced in section \ref{sec:general}.
    It will be convenient to work in coordinates where the AdS₅ metric has conformally flat slices along a chosen \textit{fifth direction} $ξ$ \ie, AdS₅ $\equiv ℝ \times_{a(ξ)²} \rm{MkW}_4$.
    \begin{equation}
    \D s² = \D ξ² + a(ξ)²η_{ab}\D x^a \D x^b,
    \end{equation}
    where $a(ξ)\coloneqq \exp(k ξ)$, $1/k=L$ is the AdS radius, indices $(a,b)\in\left\{0,1,2,3\right\}$ as before,
    $η_{ab}$ is the Minkowski metric in four dimensions and Greek letter indices are five dimensional \ie, $x^α \in (ξ,x^a)$. 
    
    Let us choose two AdS₅ spaces with radii $L_\pm = 1/k_\pm$ in the regions $ξ\gtrless 0$ respectively. In order for this composite spacetime for all $ξ\in ℝ$ to be a solution of Einstein's equations, a source needs to be present at the discontinuity at $ξ=0$. In this case, the source is a brane situated at $ξ=0$ which has a stress tensor proportional to the difference of its extrinsic curvatures as seen from either side. This, along with the uniqueness of the induced metric on the brane goes by the name of Israel-Lanczos-Sen's thin shell \textit{junction conditions} \cite{Israel:1966rt,doi:10.1002/andp.19243791403,doi:10.1002/andp.19243780505},
    \begin{equation}\label{eq:thin-shell-junction-conditions}
    \begin{split}
    \left[γ_{ab}\right]^+_- &=0,\\
    S_{ab} &= -\frac{1}{\kappa_5^2}\left(\left[K_{ab}\right]^+_--\left[K\right]^+_-γ_{ab}\right),
    \end{split}
    \end{equation}
    where the extrinsic curvature $K_{ab}$ is defined (in terms of the normal to the shell $n_α$ and tangents $e^α_a \coloneqq \D x^α/\D y^a$) as $K_{ab}=n_{α;β}e^α_a e^β_b$. In these coordinates, the extrinsic curvature becomes $K_{ab}=(1/2)∂_ξ γ_{ab}$.
    Note that the second condition is the same as the one we used in \eqref{eq:second-junction} while the first condition was automatic in section \ref{sec:general} when choosing proper time on the shell.
    
    In the absence of matter on the brane and in the limit of large proper radius ($ar→∞$), the brane can be taken to be flat and lie at $ξ=0$. However, in the presence of matter, the brane is not expected to be flat any more. The deformation of the brane in the presence of matter was discussed in \cite{Garriga:1999yh,Giddings:2000mu} and is usually referred to as \textit{brane bending}. 
    Before analyzing the effect of bending in detail for our setup, let us first look at a matter perturbation and the corresponding bending at leading order.
    
    Let us consider the perturbation of the bulk metric in the presence of a small amount of matter on the brane (resulting from the presence of strings in the bulk). Off the brane and away from the strings, there are no sources and the metric can be chosen to be transverse and traceless and is given by the junction conditions. There are two perturbation parameters in this case -- first, the matter density ($χ$) and second, the inverse of the proper radius ($ar$). The metric can be thought of as a double expansion in these parameters $χ/r$ and $1/(kar)$. Up to leading order in large proper radius, a transverse traceless perturbation in response to matter density $χ$ is given by \footnote{We will derive the metric to linear order in $χ/r$ and all orders in $(kar)^{-1}$ in section \ref{subsec:lineargrav-r} and this metric is just the large proper radius expansion of the metric in equation \eqref{pert-metric} \ie, expanded to $(kar)^0$.}
    \begin{equation}
    \label{eq:leading-unbent}
    \D s² = \D ξ² + a(ξ)²\left[- \left(1-\frac{4\chi}{r} \right)  \D t^2+ \left(1+\frac{2\chi}{r} \right)   \D r^2 +  \left(1+\frac{\chi}{r} \right) r^2 \D \Omega_2²\right].
    \end{equation} 
    The position of the brane taking into account the bending is given by $ξ=\tilde{ξ}+f(r)$. Shifting the radial coordinate by a constant $r=\tilde{r}-χ$ and dropping quadratic terms in $f$ or its derivatives gives
    \begin{equation}
    \label{eq:sch-brane}
    \D s² = \D \tilde{ξ}² + a(\tilde{ξ})²\left[- \left(1-\frac{4\chi}{r} + 2 kf(r) \right)  \D t^2+  \left(1+\frac{2\chi}{r}+2 kf(r)\right)  \D r^2 +  r^2 \D \Omega_2²\right],
    \end{equation}
    For this to be a vacuum solution of Einstein's equations at lowest order in $χ/r$ and $1/(kar)$, 
    \begin{equation}\label{eq:f3}
    kf(r)=\frac{\chi}{2r}=\frac{M}{3r} ,
    \end{equation}
    where $\chi = 2M/3$, and $M$ is the effective mass from the four dimensional point of view.
    
    Let us now do a more careful analysis, taking into account the presence of sources.
    Starting from equation \eqref{eq:thin-shell-junction-conditions} and splitting the stress tensor on the brane into the brane tension $σ$ (equation of state $ρ=-p$) and other matter $T_{ab}$ \ie, $S_{ab} = -σ γ_{ab} + T_{ab}$,
    the second junction condition becomes
    \begin{equation}\label{eq:junction}
    \frac{1}{2κ_5²}\left[∂_ξ γ_{ab}\right]^+_- +\frac{σ}{3}γ_{ab}=-\left(T_{ab}-\frac{1}{3}Tγ_{ab}\right).
    \end{equation}
    In the absence of matter on the brane \ie, $T_{ab}=0$, the brane is intrinsically flat and has a critical tension given by
    \begin{equation}\label{eq:crittension}
    σ = \frac{3}{κ_5²}\left(k_- - k_+\right).
    \end{equation}
    Now consider a perturbation of the AdS₅ metric due to the presence of some matter both on the brane and in the bulk. These perturbations can be chosen to be Gauss normal \ie, $g_{αβ} \mapsto g_{αβ}+δg_{αβ}$ such that
    \begin{equation}
    δg_{μξ}=δg_{ξξ}=0.
    \end{equation}
    This causes a change in the induced metric $γ_{ab} \mapsto γ_{ab} + \hat{h}_{ab}$, which in our coordinates becomes $a(ξ)²η_{ab} \mapsto a(ξ)²η_{ab}+ \hat{h}_{ab}$. The second junction condition then becomes (recall that this is evaluated on the shell)
    \begin{equation}\label{eq:junctionhat}
    \frac{1}{2κ₅²}\left[∂_ξ \hat{h}_{ab}\right]^+_- +\frac{σ}{3}\hat{h}_{ab}=-\left(T_{ab}-\frac{1}{3}Tη_{ab}\right),
    \end{equation}
    where $σ$ is the critical tension in equation \eqref{eq:crittension}. 
    
    In the presence of matter, the no longer expected to be at $ξ=0$. This makes it difficult to apply the junction conditions since the position of the brane would be given by two different functions $ξ=f_+(x)$ and $ξ=f_-(x)$ as seen from either side of the brane. To get around this, we can make a gauge transformation on the outside and another on the inside of the brane, following \cite{Garriga:1999yh,Giddings:2000mu}. Demanding that the new coordinates are also Gauss normal and that the brane is located at $ξ=0$ in the new coordinates, we get, following \cite{Padilla:2004mc},
    \begin{equation}\label{eq:changegauge}
    ξ\mapsto ξ-f(x), \qquad x^{a} \mapsto x^a+\frac{1}{2k}\left(1-a^{-2}\right)∂^af(x)+q^a(x),
    \end{equation}
    where $q^a(x)$ is an arbitrary function of the transverse coordinates. As mentioned above, we need to apply different coordinate transformations on either side of the brane to flatten it and so $q^a(x), f(x)$ and $k$ are different on either side of the brane. In these new coordinates, the the metric perturbation becomes
    \begin{equation}\label{eq:metricgauge}
    \hat{h}_{ab} = h_{ab} + \frac{1}{k}\left(1-a²\right)∂_a∂_bf +2kfγ_{ab}-a²\left(∂_aq_b+∂_bq_a\right),
    \end{equation}
    which when evaluated on the brane ($ξ=0\Rightarrow a=1$) becomes
    \begin{equation}\label{eq:hhtilde}
    δγ_{ab} = h_{ab}+2kfη_{ab}-2q_{[a,b]},
    \end{equation}
    where $q_{[a,b]}\coloneqq (1/2)\left(∂_a q_b+∂_b q_a\right)$ is the anti-commutator.
    Continuity of the induced metric ($δh_{ab}=0$) requires
    \begin{equation}
    \left[h_{ab}\right]^+_- =\left[kf\right]^+_- =\left[q_{[a,b]}\right]^+_- =0.
    \end{equation}
    It is practical to introduce the function $F\coloneqq kf$ since it is continuous across the shell. Using this, equation \eqref{eq:junctionhat} becomes
    \begin{equation}\label{eq:secondjunction}
    \frac{1}{2κ₅²}\left[∂_ξ h_{ab}\right]^+_- + \frac{σ}{3}h_{ab} = -Σ_{ab},
    \end{equation}
    where $Σ_{ab}$ is defined as
    \begin{equation}
    Σ_{ab} = \left(T_{ab} - \frac{1}{3}Tη_{ab}\right)-\frac{1}{κ₅²}\left(\frac{1}{k_+}-\frac{1}{k_-}\right)∂_a∂_bF,
    \end{equation}
    which gives the contribution of brane bending on the stress tensor. Therefore, choosing a bent gauge (or equivalently, considering a bent brane) modifies the stress tensor on the brane with a gauge dependent factor $F=kf$. This can be chosen so that $\Sigma_{ab}$ is traceless. This determines $F$ in terms of the trace of the stress tensor on the brane $T$ and explicitly shows how the brane bends. This is analogous to the result in \cite{Padilla:2004mc}.
    \begin{equation}\label{eq:bending}
    T=-\frac{3}{κ_5²}\left(\frac{1}{k_+}-\frac{1}{k_-}\right)\Box F.
    \end{equation}
    Note that the factor $3$ nicely cancels the $1/3$ in $kf(r)$ that we found in equation \eqref{eq:f3} by matching against Schwarzschild.
    A point mass on the brane \ie, $T_{00}\sim δ(r)>0 \Rightarrow T=-T_{00} \sim -δ(r)$, gives $\Box F \sim δ(r) \Rightarrow F \sim -1/r<0 \Rightarrow f<0$. 
    In equation \eqref{eq:changegauge}, we have defined $f(r)$ as the amount that the brane actually bends and so the coordinate $ξ$ has to be pulled down by the same amount to get a flat brane. Placing a point mass on the brane ($T_{00}\sim δ(r)>0$) implies $f(r)<0$ which means that the brane sags in response to it (case (a) of figure \ref{fig:brane_bending}). The same is true for a string that pulls the brane towards the inside (case (b) of figure \ref{fig:brane_bending}). With a string pulling on the brane from outside, the opposite is true \ie, $f(r)>0$ and the brane is pulled up as shown in case (c) of figure \ref{fig:brane_bending}.
    
    \section{Linearized gravity}
    \label{sec:lineargrav}
    
    \noindent In this section, we will first present key features of graviton propagators on our shell-world, starting from a bulk computation in momentum space. We will then repeat this computation in the position space, but now allowing for the brane to bend. This will in turn allow the possibility of having massive structures on the shell as argued before. We will finally discuss how these two apparently different computations are related via a gauge choice.
    \noindent
    
    \subsection{Linearized gravity in momentum space}
    \label{subsec:lineargrav-p}
    \noindent
    Let us consider perturbations of the AdS metric in response to sources, both in the bulk and on the brane of the form
    \begin{equation}
    \label{eq:pert-cart}
    \D s² = \D ξ² + a(ξ)²\left(η_{ab}+h_{ab}(ξ,x^a)\right)\D x^a \D x^b,
    \end{equation}
    where the perturbations are taken to be Gauss normal. Here $a(ξ)\coloneqq \exp(k ξ)$, $1/k$ being the AdS length scale.
    
    \noindent Taking the transverse coordinates, $x^a$, as standard Cartesian coordinates, the linearized Einstein equations in the world volume directions of the brane are given by \cite{Giddings:2000mu}
    \begin{equation}\label{boxh}
    \Box \bar{h}_{ab} = a^{-2}\left(-η_{ab}∂^c∂^d\bar{h}_{cd}+∂^c∂_a\bar{h}_{bc}+∂^c∂_b\bar{h}_{ac}\right)
    -\frac{1}{2}η_{ab}a^{-4}∂_ξ\left(a⁴∂_ξ\bar{h}\right)
    -2κ_5a^{-2}T_{ab},
    \end{equation}
    where $a$ is a shorthand for $a(ξ)$ , $\bar{h}_{ab}$ is the trace reversed perturbation defined as $\bar{h}_{ab} \coloneqq h_{ab} - \frac{1}{2}η_{ab}h,$
    and $h\coloneqq h_{ab}η^{ab}$ is the trace with respect to the unperturbed metric.
    Outside the source, a transverse gauge can be chosen \cite{Giddings:2000mu} \ie, $∂^a\bar{h}_{ab}=0$,
    which implies Lorenz gauge for the metric perturbations $h_{ab}$ \ie, $∂^ah_{ab}=\frac{1}{2}∂_bh$.
    
    \noindent It is easier to work with a perturbation which includes the metric factor $a(ξ)²$ (\ie, $γ_{ab} \coloneqq a(ξ)²h_{ab}$) and its trace-reversed form $\bar{γ}_{ab}=γ_{ab}-\frac{1}{2}η_{ab}γ$.
    Rewriting \eqref{boxh} in terms of $\bar{γ}_{ab}$ and taking a trace gives
    \begin{equation}\label{eq:traceequation}
    a^{-2}∂²_c\bar{γ}+3\left(∂²_ξ -4k²\right)\bar{γ} = -2κ_5²T,
    \end{equation}
    where traces are taken with respect to the background metric \ie, $\bar{γ}\coloneqq \bar{γ}_{ab}η^{ab}$ and $T\coloneqq T_{ab}η^{ab}$.
    Subtracting this from \eqref{boxh} gives an equation for the traceless piece
    \begin{equation}\label{eq:tracelessequation}
    a^{-2}∂²_cχ_{ab}+\left(∂²_ξ-4k²\right)χ_{ab} = -2κ_5²Σ_{ab},
    \end{equation}
    where $χ_{ab}$ and $Σ_{ab}$ are the traceless pieces of $\bar{γ}_{ab}$ and $T_{ab}$ respectively \ie,
    \begin{equation}
    \begin{split}
    \bar{γ}_{ab} &= χ_{ab} + \frac{1}{4}η_{ab}\bar{γ},\\
    T_{ab} &= Σ_{ab} + \frac{1}{4}η_{ab}T.
    \end{split}
    \end{equation}
    We will now try to solve for the traceless perturbation $χ_{ab}$.
    This is easier to do in momentum space. After a Fourier transformation,
    equation \eqref{eq:tracelessequation} becomes
    \begin{equation}\label{eq:pequation}
    \left(-\frac{p²}{a²}+∂_ξ²-4k²\right)\tilde{χ}_{ab}(p,ξ)=-2κ_5²\tilde{Σ}_{ab},
    \end{equation}
    where the tilde represents the Fourier transform in the transverse directions. $p^2 = -|p_0|^2+|\vec{p}|^2$ is negative for pure AdS spacetime. $p^2$  can be positive for spacelike modes with $|p_0|^2 < |\vec{p}|^2$. Such modes can appear, for example, when there is a black hole in the bulk spacetime. \footnote{Holographically, these modes are commonly interpreted as thermal modes in the boundary quantum field theory. Note, in zero temperature field theory, we can never have such modes because of spectrum condition ($ω>k$). However, when we connect the system to a heat reservoir, particles with arbitrarily large momenta can come out of the reservoir or go in keeping the total energy of the system lie within a narrow band.} In our case, with the source on the shell, we no longer have pure AdS spacetime. Let us use this fact to postulate that such non-propagating modes do exist in our case, and let us study the gravitational propagator between two bulk points across the shell for these modes. We will justify our claim through a position space computation in the next section.
    
    Equation  \eqref{eq:pequation} suggests that the structure of propagator for the modes, $\tilde{χ}_{ab}$ should be very similar to that of a minimally coupled massless scalar propagator in this spacetime. 
    In order to see this similarity explicitly, let us consider a minimally coupled massless scalar field in AdS₅ and the Green function for the scalar Laplacian in the bulk \ie,
    \begin{equation}
    \mathlarger{\Box} Δ₅\left(X^α,\tilde{X}^α\right) = \frac{δ^5(X^α-\tilde{X}^α)}{\sqrt{-G}},
    \end{equation}
    where, as before, $X^α \equiv \left\{ξ, x^c \right\}$. $α$ runs from $0$ to $4$, $c$ runs from $0$ to $3$ and $ξ$ is the fifth direction. In the background of pure AdS spacetime,
    \begin{equation}
    \D s² = \D ξ² + a²(ξ)η_{ab}\D x^a \D x^b,
    \end{equation}
    the five dimensional scalar Laplacian takes the following form
    \begin{equation}
    \begin{split}
    \mathlarger{\square} Δ₅\left(X^α,\tilde{X}^α\right) &=\frac{1}{\sqrt{-G}}∂_α\left(\sqrt{-G}g^{αβ}∂_βΔ_5\left(X^α,\tilde{X}^α\right)\right)\\
    &=\left(∂²_ξ +\frac{4a^\prime}{a}∂_ξ+\frac{∂²_c}{a²} \right)Δ₅\left(X^α,\tilde{X}^α\right).
    \end{split}
    \end{equation}
    Let us now perform the Fourier transform of the Green function $Δ_5$ in the transverse directions ($x^c$),
    \begin{equation}
    Δ₅\left(X^α,\tilde{X}^α\right) = \int \frac{d⁴p}{(2π)⁴}e^{ip(x-\tilde{x})}Δ_p(ξ,\tilde{ξ}).
    \end{equation}
    The Fourier components then satisfy
    \begin{equation}\label{eq:eq1}
    \left(∂²_ξ +\frac{4a^\prime}{a}∂_ξ-\frac{p²}{a²} \right)Δ_p(ξ,\tilde{ξ})=\frac{δ(ξ-\tilde{ξ})}{a⁴}.
    \end{equation}
    With a change of variables $\hat{Δ}_p \coloneqq a²Δ_p$, this becomes 
    \begin{equation}\label{eq:deltahat}
    \left(∂²_ξ -\frac{p²}{a²} - \underbrace{2\left(\frac{a^{\prime  2}}{a²}+\frac{a^{\prime\prime}}{a}\right)}_{=4k²} \right)\hat{Δ}_p(ξ,\tilde{ξ})=\frac{1}{a²}δ(ξ-\tilde{ξ}),
    \end{equation}
    
    \noindent The left hand side of \eqref{eq:deltahat} is exactly the same as that of \eqref{eq:pequation}. Identifying $\tilde{Σ}_{ab}$ appearing in the right hand side of \eqref{eq:pequation} as the source localized on the shell, one can, analogously, write down the scalar propagator for the traceless modes, $\tilde{χ}_{ab}(p,ξ)$. 
    \begin{equation}\label{eq:prop-equation}
    \left(-\frac{p²}{a²}+∂_ξ²-4k²\right)\Delta_{\tilde{χ}}(p; a_+,a_-)=\delta(a_+ - a_-).
    \end{equation}
    This scalar propagator carries most of the relevant features of the bulk graviton propagator in presence of a localized source on the shell. From now on, we can therefore safely treat the modes, $χ_{ab}$ as scalar modes  $\tilde{χ}$. The index structure of the actual graviton propagator can be reinstated using the exact relation between graviton and scalar propagators for localized sources \cite{{Giddings:2000mu,DHoker:1999bve}}.
    The propagator for this mode satisfies equation \eqref{eq:prop-equation}
    where the location of the shell is at $a_+ = a_- \equiv a_s$.
    Outside the source, 
    this is solved piece-wise inside and outside the shell by
    \begin{equation}\label{eq:chiinout}
    \begin{split}
    \Delta_{\tilde{χ}}^+(p;a_+,a_-) &= A(p,a_-) K_2\left(\frac{p}{a_+ k_+}\right)+B(p,a_-) I_2\left(\frac{p}{a_+ k_+}\right),\\
    \Delta_{\tilde{χ}}^-(p;a_+,a_-) &= C(p,a_+) K_2\left(\frac{p}{a_- k_-}\right),
    \end{split}
    \end{equation}
    where $K_2$ and $I_2$ are modified Bessel functions. $I_2(p/ak)$ diverges at small $a$, \ie, at the Poincaré horizon ($a→0$), so regularity at the horizon excludes it in $Δ_{\tilde{χ}}^-$.
    Continuity of $Δ_{\tilde{χ}}$ across the shell gives the first junction condition and the stress tensor on the shell determines the jump in the derivative. These are the thin shell junction conditions
    \begin{equation}
    \label{eq:junc-cond}
    \begin{split}
    \Delta_{\tilde{χ}}^-(p;a_+,a_s) &= \Delta_{\tilde{χ}}^+(p;a_s,a_-)\\
    \frac{1}{2κ₅²}\left[\frac{∂}{∂ξ_i}Δ_{\tilde{χ}}^i(p;a_+,a_-)\big|_{a_i→a_s}\right]^{i=+}_{i=-} +\frac{σ}{3}   \Delta_{\tilde{χ}}^+(p;a_s,a_-)&=\frac{1}{2κ₅²}.
    \end{split}
    \end{equation}
    Eliminating $A(p, a_-)$ and $C(p, a_+)$ from equation \eqref{eq:prop-equation} by imposing these junction conditions gives
    \begin{equation}\label{eq:chisol}
    \begin{split}
    \Delta_{\tilde{χ}}^+(p; a_+,a_-) &=  A(p,a_-) K_2\left(\frac{p}{a_+ k_+}\right)+B(p,a_-) I_2\left(\frac{p}{a_+ k_+}\right)\\
    &= -\left[\frac{1}{g_K(p,a_s)}+B(p,a_-)\frac{g_I(p,a_s)}{g_K(p,a_s)}\right]K_2\left(\frac{p}{a_+ k_+}\right) + B(p,a_-) I_2\left(\frac{p}{a_+ k_+}\right),
    \end{split}
    \end{equation}
    where the functions, $g_K(p,a_s)$ and $g_I(p,a_s)$ are defined as
    \begin{equation}\label{eq:gKgI}
    \begin{split}
    g_K(p,a_s) &\coloneqq \frac{p}{a_s K_2\left(\frac{p}{k_-a_s}\right)}\left[K_1\left(\frac{p}{k_-a_s}\right)K_2\left(\frac{p}{k_+a_s}\right)-K_1\left(\frac{p}{k_+a_s}\right) K_2\left(\frac{p}{k_-a_s}\right)\right],\\
    g_I(p,a_s) &\coloneqq \frac{p}{a_s K_2\left(\frac{p}{k_-a_s}\right)}\left[K_1\left(\frac{p}{k_-a_s}\right) I_2\left(\frac{p}{k_+a_s}\right) + I_1\left(\frac{p}{k_+a_s}\right) K_2\left(\frac{p}{k_-a_s}\right)\right].
    \end{split}
    \end{equation}
    Let us examine the behavior of the modes $\tilde \chi$ and their propagator near the boundary \ie, $a→∞$. In this limit one can expand the Bessel functions appearing in the bulk propagator, \eqref{eq:chisol}
    \begin{equation}
    \label{eq:expand-Bessel}
    \begin{split}
    \lim\limits_{a→∞} K_2\left(\frac{p}{a k}\right) &=\frac{2 a^2 k^2}{p^2}-\frac{1}{2}+\frac{4 p^2 \log (a)-4 p^2 \log \left(\frac{p}{2
    		k}\right)-4 \gamma  p^2+3 p^2}{32 a^2 k^2}+\ldots,\\
    \lim\limits_{a→∞} I_2\left(\frac{p}{a k}\right) &= \frac{1}{8}\frac{p²}{a²k²}+\ldots,
    \end{split}
    \end{equation}
    where $\gamma$ is the Euler-Mascheroni constant. Terms which vanish at the boundary are normalizable modes in the bulk and yield energy in the holographically dual theory on the boundary. On the other hand, divergent terms at the boundary are non-normalizable modes in the bulk and can be interpreted holographically as a change in the boundary theory. Since we do not expect a change of energy in the boundary theory in a nucleation event such as the creation of the brane, we need the normalizable modes to vanish. For massless scalar modes (also for the traceless modes for graviton fluctuations) in AdS$_5$, this corresponds to vanishing of terms that go as $a^{-2}$. \footnote{The log term in \eqref{eq:expand-Bessel} is a consequence of working with odd-dimensional AdS spacetime. From the holographic renormalization, such terms can be shown to be directly related to the Weyl anomaly of the dual boundary field theory \cite{deHaro:2000vlm}. This log term term can be ignored so long as the boundary metric is flat as is in our case.} This fixes the constant, $B(p,a_-)$ as
    \begin{equation}
    B(p,a_-) = -\frac{η}{η g_I(p,a_s)+g_K(p,a_s)}, \qquad \textrm{ where }\quad
    η\coloneqq 3-4γ+4\ln 2.
    \end{equation}
    With this value of $B(p,a_-)$, the bulk propagator $ \Delta_{\tilde{χ}}^+(p; a_+,a_-)$ in equation \eqref{eq:chisol}, can be evaluated on the shell in the small momentum limit to give
    \begin{equation}\label{eq:chisolfull}
    \Delta_{\tilde{χ}}^+(p; a_+,a_-) =  \frac{a_+²}{p²}\left(\frac{2k_-k_+}{k_- - k_+}\right)+\mathcal{O}\left(p^0\right)
    \Rightarrow \Delta_{\tilde{χ}}^{\rm{shell}}(p; a_s,a_s) = \frac{a_s²}{p²}\left(\frac{2k_- k_+}{k_- - k_+}\right)+\mathcal{O}\left(p^0\right).
    \end{equation}
    The leading order term is the four dimensional scalar propagator on the shell. The five dimensional bulk-bulk propagator in the Randall-Sundrum braneworld \cite{Randall:1999ee,Randall:1999vf}, as computed in \cite{Giddings:2000mu}, splits into a four dimensional propagator and a piece corresponding to the Kaluza-Klein modes. In the above result, we have expanded in small momentum and have recovered the zero mode of the five dimensional graviton corresponding to the four dimensional graviton localized to the shell. The other modes would appear as higher order corrections in $p$.                 
    
    The expression for the graviton mode on the shell corresponding to equation \eqref{eq:chisolfull}, can be obtained by convoluting the propagator with the source, $-2κ₅²\tilde\Sigma$, given in \eqref{eq:pequation}. In momentum space this is a simple product yielding
    \begin{equation}\label{eq:chi4d}
    {\tilde{\chi}}_s
    (p; a_s) = - 2κ₅² {\tilde\Sigma}\frac{a_s²}{p²}\left(\frac{2k_- k_+}{k_- - k_+}\right).
    \end{equation}
    This is the result that was derived in \cite{Banerjee:2018qey} demonstrating the appearance of four dimensional gravity with an effective Newton's constant of $G_4\coloneqq 2G_5 k_+k_-/(k_--k_+)$. 
    It is worth mentioning at this point that in this derivation we have worked in a gauge where the brane is placed at $ξ=0$. However, as we discussed in section \ref{sec:brane_bending}, the presence of matter causes the brane to bend and we must change coordinates accordingly to \textit{straighten out} the brane. We saw that this implies that matter placed directly on the brane, or caused by a string pulling from inside, contributes with a negative stress tensor. This implies that $\tilde{Σ}_{ab}<0$ and with the minus sign in equation \eqref{eq:chi4d}, four dimensional gravity is correctly reproduced.
    In fact, we can take the analysis further and verify that this, which comes from the five dimensional junction conditions, correctly reproduces the four dimensional junction conditions.
    Fourier transforming equations \eqref{eq:bending} and \eqref{eq:hhtilde} we get
    \begin{equation}\label{eq:fouriertransformT}
    \tilde{T} = \frac{3}{κ_5²}\left(\frac{1}{k_+}-\frac{1}{k_-}\right)p² \tilde{F}, \quad
    \textrm{ and } \quad 
    \tilde{h}_{ab} = \tilde{χ}_{ab} + 2\tilde{F}η_{ab}+2ip_{[a}\tilde{q}_{b]}.
    \end{equation}
    Using equation \eqref{eq:chi4d} we get
    \begin{equation}
    \begin{split}
    \tilde{h}_{ab} &= \tilde{χ}_{ab} + 2\tilde{F}η_{ab}+2ip_{[a}\tilde{q}_{b]}\\
    &= -\frac{\tilde{Σ}_{ab}}{αp²}+2\tilde{F}η_{ab}+2ip_{[a}\tilde{q}_{b]}\\
    &= -\frac{1}{αp²}\left(\left(\tilde{T}_{ab} - \frac{1}{3}\tilde{T}η_{ab}\right)+\cancel{\frac{1}{κ₅²}\left(\frac{1}{k_+}-\frac{1}{k_-}\right)p_ap_b\tilde{F}}\right)+2\tilde{F}η_{ab}+\cancel{2ip_{[a}\tilde{q}_{b]}},
    \end{split}
    \end{equation}
    where
    \begin{equation}
    α \coloneqq \frac{1}{4κ_5²}\left(\frac{1}{k_+}-\frac{1}{k_-}\right),
    \end{equation}
    and the arbitrary function $\tilde{q}_a$ is chosen so that is cancels the term proportional to  $p_ap_b\tilde{F}$ in $\tilde{Σ}_{ab}(p)$. Using \eqref{eq:fouriertransformT} this gives
    \begin{equation}\label{eq:onehalf}
    \begin{split}
    \tilde{h}_{ab}&=-\frac{1}{αp²}\left(\tilde{T}_{ab}-\frac{\tilde{T}}{3}η_{ab}\right)
    +\frac{\tilde{T}}{6αp²}η_{ab}\\
    &= -\frac{1}{αp²}\left(\tilde{T}_{ab}-\frac{\tilde{T}}{2}η_{ab}\right)
    = -\frac{2κ²_4}{p²}\left(\tilde{T}_{ab}-\frac{\tilde{T}}{2}η_{ab}\right),
    \end{split}
    \end{equation}
    where $κ²_4 \equiv 8πG_4$. This is precisely the four dimensional junction condition with the right four dimensional Planck constant. The minus sign is because of the sign of the stress tensor as discussed before.

    Let us conclude with few more comments on the main result of this section, equation \eqref{eq:chi4d}. 
    For small momentum, ${\tilde{\chi}}_s$ gives the four dimensional graviton propagator on the shell thereby producing four dimensional gravity.
    For large momentum, ${\tilde{\chi}}_s$ decays exponentially fast towards zero. It can be verified from equations \eqref{eq:chisol} and  \eqref{eq:gKgI} that the crossover in the behavior of ${\tilde{\chi}}_s$ occurs when $p\sim a_s k_+$. This is compatible with a holographic picture where the dual CFT on a shell at large and fixed $ξ$ is modified only at large distances when the bulk geodesics corresponding to correlation functions in the CFT dip down far enough into the bulk to hit the bubble wall. Short distance correlators in the boundary CFT do not see the shell and remain unmodified. 
    
    \subsection{Linearized gravity in position space}
    \label{subsec:lineargrav-r}
    So far, we have solved for the traceless perturbation in momentum space. In order to find all components of the metric perturbations in a usable form, we would like to find the corresponding perturbation in position space. To do this, let us study perturbations of the AdS$_5$ metric in response to matter on the brane which, as discussed in the previous section, can be the endpoint of a string in the bulk. These perturbations ($χ_{ab}$) can be chosen to be Gauss normal and in the Lorenz gauge. Considering spherically symmetric perturbations on the brane \footnote{contrary to the perturbation considered in \eqref{eq:pert-cart}.}, the metric (changing coordinates from $ξ$ to $z\coloneqq e^ξ$ and setting $k=1$) can be written as
    \begin{equation}\label{eq:perturbation}
    \begin{split}
    \D s²   &= \frac{\D z²}{z²} + z² η_{ab}\D x^a \D x^b + \overbrace{χ_{ab} \D x^a \D x^b}^{\rm{perturbation}} \\
    &= \frac{\D z²}{z²} + z²\left[-\left(1+h_t(r,z)\right)\D t²+\left(1+h_r(r,z)\right)\D r²+\left(1+h_a(r,z)\right)r²\D \Omega_2²\right].
    \end{split}
    \end{equation}
    Further, we make an ansatz $h_i(r,z)=z^{n-2}f_i(rz)$, where $rz$ is the proper radius. \footnote{Choosing $h_i(r,z)=z^{n-2}f_i(rz)$ makes the perturbation to the AdS metric $χ_{ab}\sim z^nf_i(rz)$ after including the $z²$ factor in front of the four dimensional part of the metric.} Making a gauge choice that the perturbation is traceless ($h_{ab}η^{ab}=0$) gives the following linearized Einstein's equation for $f_t(rz)$
    \begin{equation}\label{eq:fteq}
    \left(1+k²z²r²\right)f_t^{\prime\prime}(zr)+\left(\frac{2}{zr}+\left(2n+1\right)k²zr\right)f_t^\prime(zr)+(n²-4)k²f_t(zr)=0,
    \end{equation}
    where as usual, primes denote derivatives with respect to the argument.
    In terms of $f_t$, the radial part $f_r$ is given by a remarkably simple first order equation
    \begin{equation}
    %\begin{split}
    zrf_r^\prime(zr)+3f_r(zr)+f_t(zr) = 0 ,
    \end{equation}
    and the angular part follows from vanishing of the trace
    \begin{equation}
    f_t(zr) + f_r(zr) + 2f_a(zr) = 0.
    \end{equation}
    For $n=3$, equation \eqref{eq:fteq} has two solutions
    \begin{align}
    \label{eq:f1-f2}
    f_t^{(1)}(zr) &= \frac{1}{zr},&
    f_t^{(2)}(zr) &= \frac{3+2z²r²}{\left(1+z²r²\right)^{3/2}}.
    \end{align}
    The first solution $f_t^{(1)}(zr)$ gives a perturbation $χ_{tt}=z²h_t=z³f_t=z²/r$, while the second solution gives a perturbation which has the following asymptotics
    \begin{align}
    \lim\limits_{r \ll 1}χ_{tt}^{(2)} &\sim z³,& \lim\limits_{r \gg 1}χ_{tt}^{(2)} &\sim \frac{z²}{r}.
    \end{align}
    The perturbation $χ_{tt}^{(1)}$ behaves, in fact, like the CHR black string \cite{Chamblin:1999by} solution at short distances and is singular at $r=0$. So it needs the presence of string with a delta function source $δ(r)$. This is a string with non-constant tension \cite{Mann:2006yi} and is not related to the constant tension string we study. We discard this solution and focus on $χ_{tt}^{(2)}$.
    This gives the following radial and angular perturbations
    \begin{align}
    f_r &= -\frac{1}{\sqrt{1+r²z²}}, & f_a &=-\frac{2+r²z²}{2\left(1+r²z²\right)^{3/2}}.
    \end{align}
    So, the complete perturbed metric to first order is
    \begin{equation}
    \label{pert-metric}
    \begin{split}
    \D s_{\rm{brane}}²=&-z²\left(1+{c_1}z\frac{3+2z²r²}{\left(1+z²r²\right)^{3/2}}\right)\D t²+z² \left(1-{c_1}z\frac{1}{\sqrt{1+r²z²}}\right)\D r²\\ &+z²r²\left(1-{c_1}z\frac{2+r²z²}{2\left(1+r²z²\right)^{3/2}}\right)\D \Omega_2²+\frac{\D z²}{z²}.
    \end{split}
    \end{equation}
    This behaves like the CHR solution at large $r$ but does not need a source. The leading contribution at large distance is what we used in \eqref{eq:leading-unbent} to obtain the Schwarzschild metric on the brane, \eqref{eq:sch-brane}. Taking the brane bending into account, we find that there is an effective source, $Σ_{ab}=δ(r)/2κ_5²$, induced on the brane as discussed at the end of section \ref{sec:brane_bending}. We conclude that the $n=3$ solution corresponds to perturbations of the metric due to a point source in the brane world (possibly induced by a pulling string) at $r=0$. These grow towards the boundary of AdS and are therefore non-normalizable perturbations.
    
    Having understood the metric sourced by a mass on the brane, let us now study the metric induced by the stretched string. In the presence of a source in the bulk, the perturbations can be taken to be cylindrically symmetric around the string but do not need to be in the Lorenz gauge. They are of the general form of equation \eqref{eq:perturbation} with a non-zero perturbation in the $z$ direction. Making the ansatz $h_i(r,z)=z^{n-2}f_i(rz)$ as before, but now with $n=0$ for a string with constant tension, Einstein's equations can be solved to linear order to get the metric outside a constant tension string stretching in the bulk.
    \begin{equation}
    \begin{split}
    \D s_{\rm{brane}}²=&
    -z²\left[1-\frac{c_2}{z²}\left(3+2r²z²-\frac{2}{rz}\left(1+r²z²\right)^{3/2}\right)\right]\D t²\\
    &+z² \left[1-\frac{c_2}{z²}\left(1+2r²z²-2rz\sqrt{1+r²z²}\right)\right]\D r²\\
    &+z²r²\left[1-\frac{c_2}{3z²}\left(3+2r²z²-\frac{2}{rz}\left(1+r²z²\right)^{3/2}\right)\right]\D \Omega_2²\\
    &+\left[1 -{c_2}\left(2+\frac{2}{rz}\sqrt{1+r²z²}\right)\right]\frac{\D z²}{z²}.
    \end{split}
    \end{equation}
    
    Away from the string (at large $r$), the perturbations go as $1/z²r²$, while close to the source (at small $r$), $h_t \sim h_z \sim 1/rz$.
    
    \subsection{Final comments on the role of gauge choice}
    As promised at the beginning of this section, we will conclude with a few comments on how the computations in the last two sections are connected through a choice of gauge.
    We showed that depending on whether we choose the brane to be straight or bent, the physical interpretation would be quite different. In particular, we noted that the bent gauge automatically allows for an effective delta function source on the brane. This is what connects our computation in momentum space to that in the position space. While in the first case, we kept the position of the brane fixed, in the later we allowed brane bending. Noting that the Fourier transform of the solution $f_t^{(2)}(zr)$ in \eqref{eq:f1-f2} is nothing but the modified Bessel function $K_2$ (see appendix \ref{sec:FTK2}), we intuitively see that if we had chosen to work out the fluctuations in the momentum space, but taking into account the bending of the brane, we would have needed to work with the pure $K_2$ solution on both sides of the brane. This would amount to setting $B(p,a_-)=0$ in \eqref{eq:chiinout}. By self consistency, then, the brane is bent so that the presence of the $K_2$ does not induce any matter on the brane, such that we find a pure Schwarzschild metric. Alternatively, we can perform a coordinate transformation, shifting $z$ as a function of the coordinates on the brane, to obtain a straight brane. This will generate terms of the form $I_2$ through mixing. Again, we should make sure that there is no matter induced on the brane, which corresponds to there being no normalizable modes. This implies a specific relative coefficient of the $K_2$ and $I_2$ so that the subleading $1/z^2$ mode is canceled. This would precisely correspond to the momentum space computation that we have performed.
    
    \section{Discussion and outlook}\label{sec:conclusion}
    In this work, we have provided a clear understanding of backreaction on our shellworld due to localized and extended sources in the bulk. This complete picture, in turn, suggests several interesting directions for further studies that we are currently pursuing.
    
    \subsection{One more view from the bulk: possible model with localized gravity}
    
    We have argued that there is a consistent holographic picture of the dark brane, with massive particles realized as strings hanging from the boundary of AdS. Even though gravity is mediated by the bending of the brane, gravitational effects extend all the way to the boundary through the non-normalizable modes. Interestingly, there is an alternative possibility involving two dark branes where gravity is fully localized in the fifth direction. Physically, it would correspond to two branes about to collide. As shown in figure 2, one can imagine strings stretching between the two branes, making each of the branes bend in response. Our universe, together with the matter it contains, is supported on this combined structure. Thus, it is perhaps better to think of the two branes as one thick brane with an internal structure. As the universe expands, the sandwich gets thinner, implying a dramatic end point when the bubbles collide. As explained in \cite{Banerjee:2019fzz}, the distance between the two branes can be microscopic and still allow for plenty of proper time on the branes before collision.
    
    In this picture, any gravitational perturbation is localized to the sandwich universe and decays as we move away from it. The holographic picture corresponds to slicing the sandwich in two, assuming the two sides to be mirrors of each other, and inserting a holographic screen. It also requires that the two branes are sufficiently far away from each other. We will investigate this setup further in a forthcoming paper.
    
    \begin{figure}
    	\centering
    	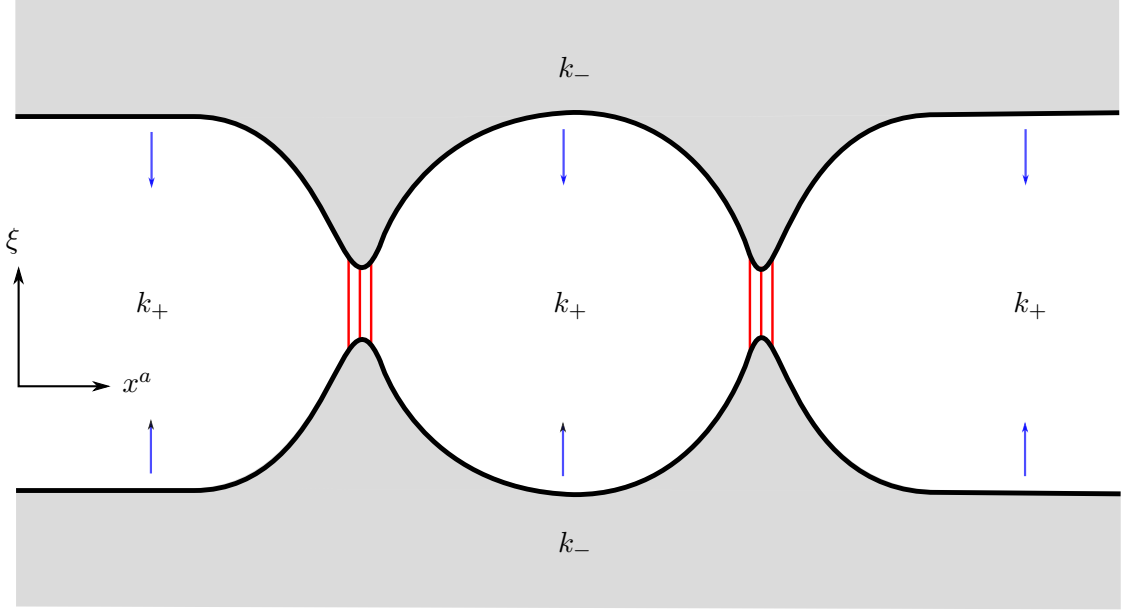
    	\caption{A cartoon representing strings (in red) stretching between two copies of the world brane (in black) that can be close together. The energy density of the strings and their endpoints on the branes causes them to bend. As the four dimensional universe within the brane expands, the branes move closer to each other (shown with blue arrows).  The gravitational perturbation from the strings is localized to the branes and the false vacuum in between. The perturbations decays away as we move out from the sandwich universe into the true vacuum.}
    \end{figure}
    
    \subsection{Black objects on the brane}
    
    An important challenge is to construct solutions corresponding to black holes. Given that massive particles corresponds to stretched strings, it seems natural to consider such objects in the context of the sandwich universe discussed above.  Black holes should form if, for instance, a shell of stretched strings such as in figure 2, were to collide and approach their Schwarzschild radius. One possible outcome would be a five dimensional black hole acting as a connection between the two braneworlds. Inspired by \cite{Danielsson:2017riq,Danielsson:2017pvl}, we believe there are also other possibilities that should be investigated. In particular, the collision of the strings could initiate the nucleation of a bubble of true vacuum that could prevent the formation of a black hole. The idea would be that such a structure could remain stable due to Unruh radiation in parallel to what was proposed in \cite{Danielsson:2017riq,Danielsson:2017pvl}. We will provide details of such constructions in an upcoming publication.

    \subsection{Holography on a cut off brane}
    The derivation of the bulk propagator in momentum space presented in section \ref{subsec:lineargrav-p} shows that the effective four-dimensional gravity on our shellworld is perfectly consistent with the expectations from holography. 
    By holography we here mean the description of gravitational physics in the bulk AdS  in terms of the field theory living on its boundary at $a = \infty$. Here, it is worth mentioning several previous attempts towards investigating the gravitational duals of collapsing shells in asymptotically AdS spacetime \cite{Danielsson:1998wt, Danielsson:1999zt,Danielsson:1999fa, Giddings:2001ii}. However, with a view to understand the gravitational dual of the dynamics of thermalization in the boundary quantum field theory, all of these studies were centered on collapsing shells. It would nevertheless be straightforward to do a similar computation of boundary propagators in our case. Although our model of expanding bubble is physically very different from the scenario of collapsing shells in AdS, the holographic techniques we need to employ in order to compute the boundary correlators would be quite similar. In particular, taking the backreaction into account, it would be worth investigating the signatures of a possible formation on black objects on the shell from the perspective of the boundary field theory \cite{Gregory:2008br}. However, it would perhaps be  even more interesting and illuminating to understand the whole picture in terms of the the holography on the shell. Considering the shell to be a finite radial cut off from the perspective of both the AdS spacetimes, one possible way to understand this holographically is to consider a CFT deformed by an irrelevant operator, typically termed as $T\bar T$ deformation in literature \cite{McGough:2016lol,Taylor:2018xcy,Hartman:2018tkw}. This is a very active field of research in the holography community and would be instrumental in connecting the five-dimensional bulk physics in our set up to a deformed CFT on the shellworld. We hope to report on this soon.
    \section*{Acknowledgments}
    We would like thank Giuseppe Dibitetto, Johanna Erdmenger, Miguel Montero, Lisa Randall, Marjorie Schillo, Cumrun Vafa and Irene Valenzuela for several stimulating discussions. We would also like to thank the referees for their valuable comments. The work of S.B. is supported by the Alexander von Humboldt postdoctoral fellowship.
    \appendix
    \section{Fourier transform of $K_2$}\label{sec:FTK2}
    The 3 dimensional Fourier transform of the modified Bessel function $K₂$ in spherical coordinates is given by
    \begin{equation}\label{eq:FTK2}
    \begin{split}
    \mathcal{F}(K₂(p))=\int \D ^3 \vec{p} e^{i\vec{p}\cdot \vec{r}}K_2(p)
    &= 2π \int\limits_{0}^{π}\D θ \sin θ\int \D p e^{ipr\cos θ} p² K₂(p)\\
    &= \frac{4π}{r} \int \D p p \sin pr K₂(p)\\
    &= -\frac{4π}{r} \frac{∂}{∂r} \int \D p \cos pr K₂(p)
    \end{split}
    \end{equation}
    $K_2(p)$ has an integral representation given by \cite{NIST:DLMF}
    \begin{equation}
    K_2(p) = -\int\limits_{-∞}^{∞}\D r\cos pr\frac{2r²+1}{2\sqrt{1+r²}},
    \end{equation}
    which can be inverted to give~\footnote{The Fourier cosine transform and its inverse are defined with normalization $F(ω)=(1/2)\int\limits_{-∞}^{∞}f(t)\cos ωt \D t$ and $f(t)=(1/π)\int\limits_{-∞}^{∞}F(ω)\cos ωt \D ω$ respectively.}
    \begin{equation}
    \int\limits_{-∞}^{∞}\D p \cos pr K₂(p) = -π\frac{2r²+1}{\sqrt{1+r²}},
    \end{equation}
    Using this in \eqref{eq:FTK2} gives
    \begin{equation}
    %\begin{split}
    \mathcal{F}(K₂(p)) = \frac{4π}{r}\frac{∂}{∂r}\left(\frac{2r²+1}{2π\sqrt{1+r²}}\right)= 4π² \frac{2r²+3}{\left(1+r²\right)^{3/2}}.
    %\end{split}
    \end{equation}

    % References
    \small
    \bibliography{references}
    \bibliographystyle{utphysmodb}
\end{document}